\documentclass[fleqn,usenatbib]{mnras}

\usepackage{newtxtext,newtxmath}

\usepackage[T1]{fontenc}
\usepackage{natbib}

\DeclareRobustCommand{\VAN}[3]{#2}
\let\VANthebibliography\thebibliography
\def\thebibliography{\DeclareRobustCommand{\VAN}[3]{##3}\VANthebibliography}

\usepackage{xcolor}

\usepackage{graphicx}	

\usepackage{lineno}

\definecolor{dkblue}{RGB}{54, 86, 169}

\title[Searching for continuous Gravitational Waves ]{
Searching for continuous Gravitational Waves in the second data release of  the International Pulsar Timing Array
}

\author[IPTA Collaboration]{
	\parbox{\textwidth}{
		\Large
		M. Falxa,$^{1}$ \thanks{E-mail: falxa@apc.in2p3.fr}
		S. Babak,$^{1}$
		P. T. Baker,$^{2}$
		B. B\'{e}csy,$^{3}$
		A. Chalumeau,$^{4}$
		S. Chen,$^{5}$
		Z. Chen,$^{6}$
		N. J. Cornish,$^{3}$
		L. Guillemot,$^{7,8}$
		J. S. Hazboun,$^{9}$
% ...
		\begin{center} (IPTA Collaboration) \end{center}
	}
	\vspace{0.4cm}
	\\
	\emph{\normalsize \it Affiliations are listed at the end of the paper}
	}

\date{Accepted XXX. Received YYY; in original form ZZZ}

\pubyear{2021}

\begin{document}
	\label{firstpage}
	\pagerange{\pageref{firstpage}--\pageref{lastpage}}
	\maketitle

\begin{abstract}
The International Pulsar Timing Array 2nd data release is the combination of datasets from worldwide collaborations. In this study, we search for continuous waves: gravitational wave signals produced by individual supermassive black hole binaries in the local universe. We consider binaries on circular orbits  and neglect the evolution of orbital frequency over the observational span. We find no evidence for such signals and set sky averaged 95\% upper limits on their amplitude $h_{95}$. The most sensitive frequency is 10nHz with $h_{95} = 9.1 \times 10^{-15}$. We achieved the best upper limit to date at low and high frequencies of the PTA band thanks to improved effective cadence of  observations. In our analysis, we have taken into account the recently discovered common red noise process, which has an impact at low frequencies. We also find that the peculiar noise features present in some pulsars data must be taken into account to reduce the false alarm. We show that using custom noise models is essential in searching for continuous gravitational wave signals and setting the upper limit. 
\end{abstract}
\begin{keywords}
	gravitational waves -- methods:data analysis -- pulsars:general
\end{keywords}

\section{Introduction}

The goal of the Pulsar Timing Array (PTA) collaborations is to detect gravitational wave (GW) signals in the nanohertz band, where we expect to see a gravitational wave background (GWB) produced by the superposition of GW signals from the population of supermassive black hole binaries (SMBHBs) \citep{Jaffe:2002rt, Sesana:2008mz, Maiorano_2021}. Some individual SMBHBs might be brighter than most and stand above the stochastic signal; those are individually resolvable sources \citep{Sesana:2008xk, Rosado_2015}. The binaries detectable in the PTA band
are in the orbits with the period from a few months to a few years and emit almost monochromatic GWs continuously during decades; we refer to those signals as continuous GWs (CGWs) \citep{Ellis_2012, epta_cw, nanograv_cw_11yr, corbin_cornish}.

The GWs affect propagation of the  radio emission from millisecond pulsars leaving an imprint in the time of arrival (TOA) 
of pulses observed with the radio telescopes. CGWs impact TOAs from all observed millisecond pulsars in a deterministic manner characterized by parameters of the SMBHBs. In this work, we consider the data combined by the International Pulsar Timing Array (IPTA). IPTA is a consortium of NANOGRAV \citep{2015_NG}, European Pulsar Timing Array (EPTA) \citep{2016_EPTA}, Australian (PPTA) \citep{2013_PPTA} and Indian Pulsar Timing Array (InPTA) \citep{inpta_dr1} collaborations. In particular, we analyze the second data released by IPTA (IPTA DR2)
described in details in \citet{ipta_second_dr}. 

Recently, PTA collaborations have reported on the discovery of the common red noise signal, that is the stochastic signal with the spectral shape common to all pulsars in the array. Its high statistical significance was demonstrated independently by three collaborations (\cite{2020_NG_CRN, Chen_2021, Goncharov_2021}) and, with even higher statistical confidence, was assessed using the IPTA DR2 \citep{ipta_antoniadis_2022}. We do not yet know the nature of this process, and its interpretation as GW background is inconclusive: the data is not informative enough to resolve the Helllings-Downs spatial correlations \citep{hellings_downs}, which should be present in the case of the GW signal. 

In this work, we search for continuous GWs which could be present in the data in addition to the stochastic GWB. Following the steps of previous studies \citep{epta_cw, nanograv_cw_11yr, zhu_2014, B_csy_2022, B_csy_2020, nanograv_12_5_yr}, we search for a single GW signal produced by a SMBHB binary in a circular orbit.   In our study, we neglect the pulsar terms during the search and setting an upper limit on GW amplitude. However, we do an in-depth analysis of the (weak) candidate events identified as plausible GW signals. In the followup analysis on the restricted parameter space (frequency and sky position), we extend our model to include (i) pulsar term, (ii) eccentricity, (iii) extend the model beyond the assumption of a single source.   For the first time, we have included in the analysis the common red component as a part of the total noise model. 

The main results of the paper can be summarized as follows. We did not detect any CGW signal and set an upper limit on GW amplitude. We have found that the noise model plays a crucial role in interpreting PTA observations. The detailed analysis performed on the most promising candidate event
revealed that it could be explained by a time-correlated high-frequency noise in one of the pulsars. 

The paper is organized as follows. In the next Section, we will briefly describe the IPTA DR2 dataset and the data model used in the analysis. Most of the material needed for this Section is available in the literature, and we heavily refer to it, keeping only parts which are necessary for further presentation. In Section~\ref{sec:methods} we describe the methodology which we have followed to get our results presented in Section~\ref{sec:results}. In Section~\ref{sec:noise_effect} we give a detailed follow-up study of a most promising GW candidate event and demonstrate the importance of noise modelling at high frequencies. We conclude with Section~\ref{sec:concl}. Throughout the paper, we adopt geometrical units $G=c=1$.   

\section{IPTA Data Release 2  and the data model}

\subsection{IPTA DR2 dataset}

The IPTA DR2 consists of 65 stable MSPs with the duration of observations up to 30 years \citep{ipta_antoniadis_2022, ipta_second_dr} \footnote{We used only the 53 pulsars of \cite{ipta_antoniadis_2022} for our analyses.}. It combines the pulsar timing data acquired by European Pulsar Timing Array \citep{epta_dr1}, North American Nanohertz Observatory for Gravitational Waves \citep{nanograv_9yr}, and the Parkes Pulsar Timing Array \citep{ppta_dr1}. The combined data is superior to the datasets of each collaboration: (i) it has better sky coverage providing better localization of GW signals, (ii) allows better decoupling and identification of noise components due to increased number of observing backends, and (iii) reduces the number of gaps in the data due to absence of observations. We have already observed the improvement in the detection of the common red noise process in \cite{ipta_antoniadis_2022} by using IPTA data.

The combined dataset was analyzed to extract the properties of individual pulsars  (pulse frequency, spin-down, parallax, etc.) by fitting a timing model that predicts the TOAs \citep{tempo2}. Differences in predicted TOAs and measured TOAs in the dataset form the timing residuals. The residuals are the result of various noise processes as well as the interaction of the radio emission with GWs, which is the main subject of this work.

\subsection{Noise model}

The noise of each pulsar data is modelled using the Gaussian process and split into several components 
(see  \cite{ipta_antoniadis_2022} for details). The white noise (WN), that estimates the TOA measurement errors, quantifies the radiometer noise in the receiver backend system and models the jitter noise which is intrinsic to the pulsar (statistics of
pulse-to-pulse variation). The timing model (TM) corrects deterministic TOA perturbations of physical origin \citep{tempo2}. Even though we fit the TOAs for the timing model before we start the analysis, the fit might not be perfect and leave behind some residuals which we assume to be small and use a linear model\footnote{Linear in deviations from already determined parameters} to describe TM-generated errors. The low-frequency part of the data is strongly affected by red noise, which is a time-correlated process which power spectral density ($P(f)$) we describe as a simple power-law. We distinguish achromatic red noise (RN, $P^{a}_{RN}$) intrinsic to each pulsar (denoted by a subscript $a$)  due to stochastic variations in the rotation of a pulsar and chromatic (i.e. dependent on the radio frequency at which pulses are observed) dispersion measurements variations (DM, $P^{a}_{DM}$) noise caused by time-varying interstellar medium properties along the line of sight.  Those two processes are described as 

\begin{equation}
	P_{RN}(f) = \frac{A_{RN}^2}{12 \pi^2} f_{yr} ^{-3} \bigg(\frac{f}{f_{yr}}\bigg)^{-\gamma_{RN}},
\end{equation}

\begin{equation}
	P_{DM}(f) = \frac{A_{DM}^2}{12 \pi^2} f_{yr} ^{-3} \bigg(\frac{f}{f_{yr}}\bigg)^{-\gamma_{DM}} \bigg(\frac{1400MHz}{\nu}\bigg)^2,
\end{equation}
where $\nu$ is radio observation frequency and $f_{yr} = \rm{yr}^{-1}$  (see \cite{ipta_antoniadis_2022} for details).  Note that $\{A_{RN/DM}, \gamma_{RN/DM}\}$ are individual 
for each pulsar and we omit the pulsar index $\alpha$ to ease the notations. 
In addition, for the first time, we will also add the common red noise which is firmly established recently
  \citep{ipta_antoniadis_2022, nanograv_crn, Chen_2021, Goncharov_2021}, we denote it as $\{A_{\rm{crn}}, \gamma_{\rm{crn}}\}$. 
  This is a red noise component with the spectral properties shared across all pulsars in the array. The nature of this signal is still unclear, there is not enough evidence to support its GW origin, and it is a subject of current active investigations; for now, we call it "crn".  

All observations are translated to the solar system barycenter (SSB) frame. The transformation from the Earth's based frame to SSB relies on a precise knowledge of the solar system ephemeris (SSE): in this work, we use DE438 ephemeris \citep{folkner2018planetary}. It was noted that there could be unaccounted systematic errors in the SSE, which could be mistaken for a stochastic GW signal. We have included BAYESEPHEM \citep{bayesephem} in the data model to mitigate those potential errors. BAYESEPHEM is a phenomenological model that varies the orbital elements of major external planets and takes into account possible systematics in SSE; note that it might also absorb part of the stochastic GW signal.    

The GW background would require including the Hellings-Downs correlations between pulsars in the data model. We do neglect cross-correlation terms in our analysis, reducing in
practice the GW background to the detected CRN. This is justified because:
(i) any cross-correlation present in the data is rather weak; otherwise,
it would have been detected (ii) the auto-correlation part is captured already in
the CRN process that we include in the model.

All in all, the model of the timing residuals is a superposition of the noise components described above. 
In addition, we will assume (and test this hypothesis) that the data contains a deterministic continuous GW signal $s_{CGW}(t)$:

\begin{equation}
	\vec{\delta t} = {\bf{M}} \vec{\epsilon} + n_{WN} + n_{RN} + n_{CRN} + n_{DM} + s_{SSE}(t)  + s_{CGW}(t).
\end{equation}
The TM contribution is described by ${\bf{M}} \vec{\epsilon}$ where $\bf{M}$ is a design matrix and $ \vec{\epsilon} $ are the linear corrections to the timing model parameters;   $n_{WN}$, $n_{RN}$, $n_{CRN}$,  $n_{DM}$ are the components described above and correspond to  the white noise,  the individual red noise,  the common red noise and the dispersion measurement  variations noise; the signal 
$s_{SSE}(t)$ denotes the BAYESEPHEM model for SSE systematics.

We base our analysis on the noise model derived for each pulsar independently. This approximation assumes that the GW signal contribution is sub-dominant and neglected in modelling each pulsar data. In fact, the GW signal will be absorbed into the RN during this step and should be decoupled when we analyze the full array allowing RN parameters to vary. For the main analysis, we fix parameters of the WN component to their maximum likelihood values obtained from the single pulsar noise analysis. It was shown \citep{ipta_antoniadis_2022, Chalumeau_2021} that this does not affect the result for the stochastic GW signal search, and we assume the same for continuous GW search. This assumption tremendously reduces the parameter space, which otherwise would be computationally intractable. 

We model  each noise  component as a Gaussian process \citep{van_Haasteren_2014} using $\sin$ and $\cos$  as basis functions evaluated at $f_k = k/T_{obs}$ Fourier frequencies, where $T_{obs}$ is time span of observations:

\begin{equation}
	n(t) = \sum _k^N X_k \cos(2 \pi t f_k) + Y_k\sin(2 \pi t f_k),
\end{equation}
where the $X_l$ and $Y_l$ are the Gaussian distributed weights with the covariance matrix defined by the power spectral density of the noise. In our approach, we marginalize over the weights. In previous PTA analysis \citep{epta_cw, nanograv_cw_11yr, zhu_2014},
the number of Fourier components  ($N$) was fixed to  30 for both RN and DM\footnote{Note, NG have used a different model for DM, namely DMX, which is not decomposed in Fourier basis functions \citep{Lam_2016, 2015_NG}}. This choice was motivated by computational savings and that those noise components mainly contribute at the low-frequency end of PTA sensitivity. However, the recent study based on the Bayesian model selection has shown that the noise models with specific values of $N$ for each noise component and each pulsar are better supported by the data \citep{Chalumeau_2021}.

\subsection{Continuous Gravitational Waves}
\label{sec:cw}

The concept of detecting GWs with PTA was formulated in ~\cite{1978SvA....22...36S, 1979ApJ...234.1100D} . The response to a deterministic GW signal can be written as

\begin{equation}
	\label{eqn:int_cw_resid}
	s_a(t, \omega) = \int _0 ^t \frac{1}{2} \frac{\hat{p}_a^i \hat{p}_a^j}{1 + \hat{p}_a \cdot \hat{\Omega}} \Delta h_{ij} (t') dt' ,
\end{equation}
where $\hat{p}_a$ is the unit vector pointing to the pulsar $a$ in the sky, $\hat{\Omega}$ is the direction of GW propagation and  $h_{ij}$ is the GW strain in the transverse-traceless gauge ($i$ and $j$ are the spatial indices). The response depends on the GW strain at two instances of time: the time of emission of electromagnetic signal and the time of its reception:

\begin{equation}
	\Delta h_{ij} (t) = h_{ij} (t - \tau _a) - h_{ij} (t),
\end{equation}
where
\begin{equation}
	\tau _a = L_a (1 + \hat{\Omega} \cdot \hat{p}_a)
\end{equation}
and $L_a$ is the distance to the pulsar $a$.  The time difference in the strain corresponds to the light travel time between the Earth and  the pulsar with a geometrical factor.  The corresponding two terms in the expression of the timing residuals are usually referred to as Earth  $s_e$ and pulsar $s_p$ terms:
\begin{equation}
	s_a (t) = s_{p, a} (t - \tau_a) - s_{e, a} (t).
\end{equation}


The strain amplitude of a GW produced by a circular binary system is given by:

\begin{equation}
	h_{ij} (t, \Omega) = \sum _{A=+, \times} e^A _{ij} (\hat{\Omega}) h_A (t),
\end{equation}
where $e^{+,\times} _{ij}$  are two  GW polarization tensor  defined as 
\begin{eqnarray}
	e^+ _{ij} (\hat{\Omega}) &=& \hat{m}_i \hat{m}_j - \hat{n}_i \hat{n}_j ,\\
	e^+ _{ij} (\hat{\Omega}) &=& \hat{m}_i \hat{m}_j - \hat{n}_i \hat{n}_j ,\\
	\hat{\Omega} &=& -\sin \theta \cos \phi \hat{x} - \sin \theta \sin \phi \hat{y} - \cos \theta \hat{z}
\end{eqnarray}
and the unit vectors are 
\begin{eqnarray}
	\hat{m} &=& -\sin{\phi} \hat{x} + \cos{\phi} \hat{y}, \\
	\hat{n} &=& -\cos \theta \cos \phi \hat{x} - \cos \theta \sin \phi \hat{y} + \sin \theta \hat{z}, 
\end{eqnarray}
where ($\theta$, $\phi$) are the polar coordinates of the GW source sky location.


Plugging these expressions in equation (\ref{eqn:int_cw_resid}), we obtain the timing residuals expected in the PTA data for a CGW  signal coming from a circular SMBHB:

\begin{equation}
	\label{eqn:resid}
	s_a(t, \hat{\Omega}) = \sum _A F^A(\hat{\Omega})[s_A (t) - s_A (t - \tau_a)]
\end{equation}

with :

\begin{eqnarray}
		s_+(t) = \frac{\mathcal{M}^{5/3}}{d_L \omega(t)^{1/3}} &\left[ -\sin[2\Phi (t)](1+\cos^2 \iota) \cos 2\psi \right. \nonumber \\
		&\left.- 2\cos[2\Phi(t)]\cos \iota \sin 2\psi\right] ,\\
		s_\times(t) = \frac{\mathcal{M}^{5/3}}{d_L \omega(t)^{1/3}} &\left[ -\sin[2\Phi (t)](1+\cos^2 \iota) \cos 2\psi \right. \nonumber \\
		&\left. + 2\cos[2\Phi(t)]\cos \iota \sin 2\psi\right],
\end{eqnarray}

where $\mathcal{M}$ is the chirp mass, $d_L$ the luminosity distance, $\omega(t)$ the CGW orbital angular frequency,  $\iota$ is the orbital inclination to the line of sight, $\psi$ is a polarization angle and $\Phi(t)$ is the phase of CGW.

The $F^A$ are the antenna pattern functions \citep{Babak:2011mr, Sesana_2010, Ellis_2012, Taylor_2016} given as

\begin{eqnarray}
	F^+ (\hat{\Omega}) &=& \frac{1}{2} \frac{(\hat{m} \cdot \hat{p})^2 - (\hat{n} \cdot \hat{p})^2}{1 + \hat{\Omega} \cdot \hat{p}}, \\
	F^\times (\hat{\Omega}) &=& \frac{(\hat{m} \cdot \hat{p})(\hat{n} \cdot \hat{p})}{1 + \hat{\Omega} \cdot \hat{p}}.
\end{eqnarray}

In this work, we neglect the pulsar term considering it as a part of the noise, assuming that the source has evolved sufficiently over $\tau_{\alpha}$ to move the pulsar term off the earth-term frequency. Including pulsar term should improve the parameter estimation but comes with a huge price of the increase in the complexity of the likelihood surface and the dimensionality of parameter space (2 additional parameters per pulsar for phase and frequency of pulsar term, e.g. see \cite{corbin_cornish}). We foresee the possibility of following up the candidate events (identified using the earth term only) with the extended signal model (pulsar term, eccentric orbit) on the reduced parameter space. We also neglect the evolution of the GW frequency ($\omega_o = 2\pi f_{gw}$) over the observation time. The frequency evolution becomes potentially measurable for the heavy sources emitting at frequency $\ge 10^{-7}$Hz,  neglecting the frequency evolution does not prevent us from detecting the sources but introduces a bias in the measured GW frequency (overestimating it), for more details see conclusion in \cite{petito}.  So the CGW phase takes a very simple form:
\begin{equation}
	\Phi (t) = \omega _o t + \phi_0 / 2,
\end{equation}
where $\phi_0$ is  initial orbital phase. Finally, the CGW amplitude $h$ is a function of $\mathcal{M}$, $d_L$ and $f_{gw}$ given by
\begin{equation}
	h = \frac{2\mathcal{M}^{5/3} (\pi f_{gw})^{-2/3}}{d_L}.
\end{equation}

We consider the model containing only one CGW signal. This model still detects multiple CGW if they are present in the data at the non-overlapping Fourier frequencies (see \cite{Babak:2011mr} for discussion). If we find more than one candidate with sufficient statistical significance as potential GW sources, we will conduct additional investigations extending our model to include several CGWs. 
We start the analysis with 1 CGW source characterized by 7 parameters summarized in table \ref{tab:cw_params} together with their prior range (we always assume a uniform prior\footnote{For setting an upper limit we use uniform prior on the amplitude of GW strain}).

\begin{table}
	\centering
	\begin{tabular}{c|c}
		CGW parameter & Range \\
		\hline
		$\log_{10} h$ & [-18, -11] \\
		$f_{gw}$ (Hz)& [$10^{-9}$, $10^{-7}$]\\
		$\phi_0$ & [0, 2$\pi$]\\
		$\cos \iota$ & [-1, 1] \\
		$\psi$ & [0, $\pi$]\\
		$\theta$ & [0, $\pi$]\\
		$\phi$ & [0, 2$\pi$]
	\end{tabular}
	\caption{List of the CGW parameters as defined in our model with their respective ranges.}
	\label{tab:cw_params}
\end{table}

\section{Method}
\label{sec:methods}

We work within the Bayesian framework and start with running the search for the CGW signal. As mentioned 
above, we sample parameters of CGW together with the noise parameters for RN and DM. We keep the white noise parameters fixed and marginalize over the timing model and BAYESEPHEM parameters. We made two runs: with and without CRN, to check how much it affects our result. 
We always use Markov chain Monte-Carlo (MCMC) sampler \citep{ptmcmc}, and we use  ENTERPRISE \citep{enterprise}  software to construct the models and compute the likelihood and prior probability. 

We extensively use the single pulsar noise explorations runs performed before the main analysis. We have converted posteriors for the RN and DM into 2D histograms ($\{A_{RN,DM}, \gamma_{RN,DM}\}$) and use them as one of a proposal for those parameters.  This empirical proposal improves the efficiency of MCMC and reduces the autocorrelation length of the chain \citep{nanograv_cw_11yr}.

During the search, we compute the Bayes factor (BF), comparing the null model (noise only) against the model where we have a CGW signal on the logarithmically spaced frequencies. In the absence of the detection, we proceed to setting an upper limit, building a 95\% sensitivity curve. During the upper limit analysis, we used a uniform prior on the amplitude of the GW signal. 

\begin{figure}
	\centering
	\includegraphics[width=0.5\textwidth]{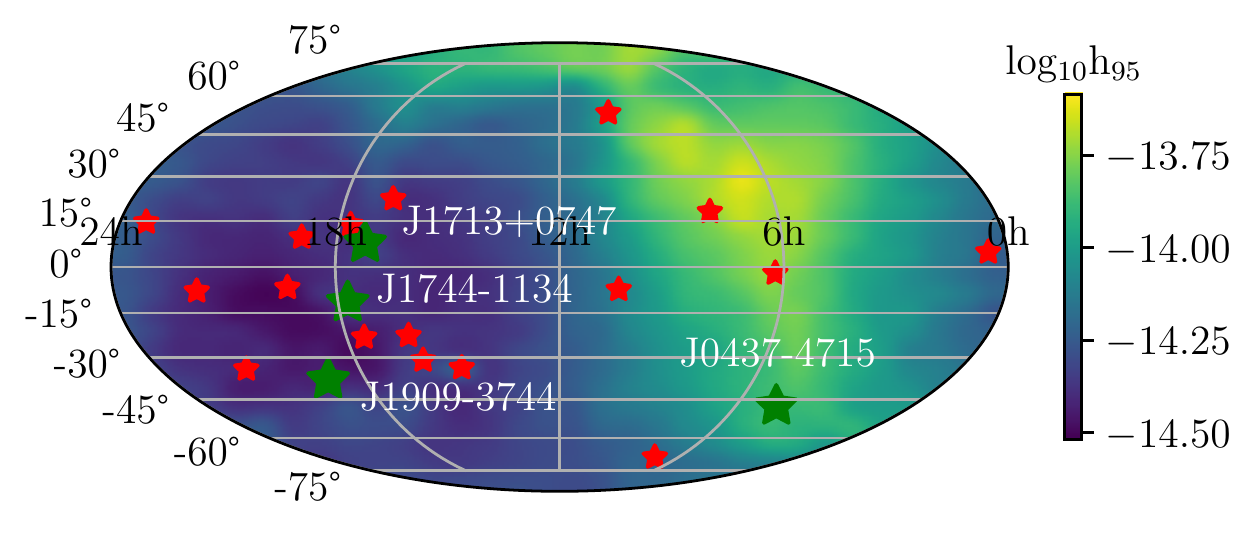}
	\caption{2D sky sensitivity of IPTA for CGW signal around most sensitive frequency. The figure was obtained computing the 95\% upper limit for $h$ on 12 patches across the sky. Gaussian interpolation was used to smoothen it across boundaries.}
	\label{fig:2d_sky}
\end{figure}

IPTA data contains 53-millisecond pulsars; however, not all of them are equally sensitive to the CGW. We have selected 21 pulsars which, on average, recover 95\% of the array's total signal-to-noise ratio (SNR) to CGW. The selected pulsars are depicted on the projected sky map in Figure~\ref{fig:2d_sky} as red stars; we have used large green stars and annotation for the four best timers. The ranking method is briefly outlined in \cite{epta_cw} and in greater detail in  \cite{epta_ranking}. This significantly reduces the computational cost without much affecting the final results \footnote{We have analyzed 21 pulsar, the noise model for each pulsar is characterized by 4 parameters  (an amplitude and a spectral index for RN and DM); Bayesphem adds 11 parameters. Finally, we have 7 parameters for the continuous wave and 2 for the common red noise, giving a total of 104 parameters to sample.}


Historically we have performed the search with the noise model with the uniform settings across all pulsars in the array; namely, we have used 30 frequency bins for modelling both RN and DM processes (as it was done in preceding work \cite{2014ApJ...794..141A, nanograv_cw_11yr}). Using this uniform setting we have obtained quite erroneous results, and after a long investigation, we realised that this noise model does not adequately describe the observational data (see section \ref{sec:noise_effect}). We have switched to 
another noise model where we have used the custom-made noise model for the six best EPTA pulsars (see  \cite{Chalumeau_2021} for details), and for other pulsars, we changed the number of used frequency bins: {\it RN30DM100} -- 30 bins for the RN and 100 bins for DM modelling. We will continue using this short-hand notation for the noise model, showing explicitly the number of  Fourier frequencies used by the base functions in the Gaussian process describing the corresponding noise. The results presented in the 
next section were obtained using the "custom" made noise model; we postpone the detailed discussion on the noise model selection and influence until Section~\ref{sec:noise_effect}.

\subsection{Model selection}
During the search, we consider two competing models: noise only and noise with 1 CGW signal. 
We compute the Bayes factor (BF) to measure which model is preferred by the IPTA DR2 dataset. 
In particular we employ hyper-model jumps to compute BF following the methods outlined in \cite{hee_hypermodel} and implemented in 
the \texttt{enterprise extensions} \citep{enterprise_extensions}.  In this method, we introduce a hyper-parameter which indexes the models; then, we perform sampling inside each model and in this parameter. The BF then is given by the ratio of the number of iterations that the chain spends in each model. For example, if we consider two models $\mathcal{M}_A$ and $\mathcal{M}_B$ with the hyper-parameter $n$, then the BF is the ratio 
\begin{equation}
	\mathcal{B}_B ^A = \frac{n_A}{n_B}
\end{equation} 
where $n_A, n_B$ counts the number of samples in the chain corresponding to the models $A, B$.  In our previous investigations, we have found that this method gives quite a reliable result, e.g. see \cite{Chalumeau_2021} where comparison is done against the evidence computation obtained with \texttt{Dynesty} nested sampling \citep{Speagle:2019ivv, Skilling:2006gxv}.

Later in the paper we will compute BF between the models where we also vary the noise.

\subsection{Upper limit} 
For obtaining the upper limit, we again use MCMC  assuming a uniform prior distribution for $\log h$ within the bounds $[-18, -11]$ on the fixed set of $f_{gw}$. We use marginalized posterior probability distribution for the CGW amplitude  $p(h)$ to set 95\% upper limit $h_{95}$ defined as :

\begin{equation}
	0.95 = \int _0 ^{h_{95}} p(h)dh.
\end{equation}
We use a grid of 100 logarithmically spaced frequencies between $10^{-9}$ and $10^{-7}$ Hz.
 The lowest bound determined by the  IPTA observational span $1/T_{obs}$ that is  $\approx 10^{-9}$ Hz while the upper bound, $10^{-7}$ Hz,  is constrained by  reduction in the sensitivity due to response ($\propto  f^{-1/3}$)  and  by our assumption that GW frequency does not evolve. We should take into account the frequency evolution of the GW signal above $10^{-7}$ Hz during the analysis, as discussed in \cite{petito}. As mentioned above, we compute the upper limit for two models of noise; with and without CRN. The inclusion of CRN  adds two more parameters to the model (an amplitude parameter $A_{CRN}$ and a spectral index $\gamma _{CRN}$), which we sample together with parameters of CGW and other noise components.

\section{Results}
\label{sec:results}


We compute BF between two models $\mathcal{B}_{custom} ^{custom+CGW}$ for a noise only model $\mathcal{M}_{custom}$ and noise+CGW signal model $\mathcal{M}_{custom+CGW}$. The subscript "custom" corresponds to the noise model we have used for these main results and distinguishes it from other noise models considered in the next section. 
Similarly to the upper limit run, we have used a log-uniform prior on $h$ bounded by [-18, -11], and the BF was computed on the grid of 100 CGW frequencies ($f_{gw}$) between 1 and 100 nHz.

\begin{figure}
	\centering
	\includegraphics[width=0.5\textwidth]{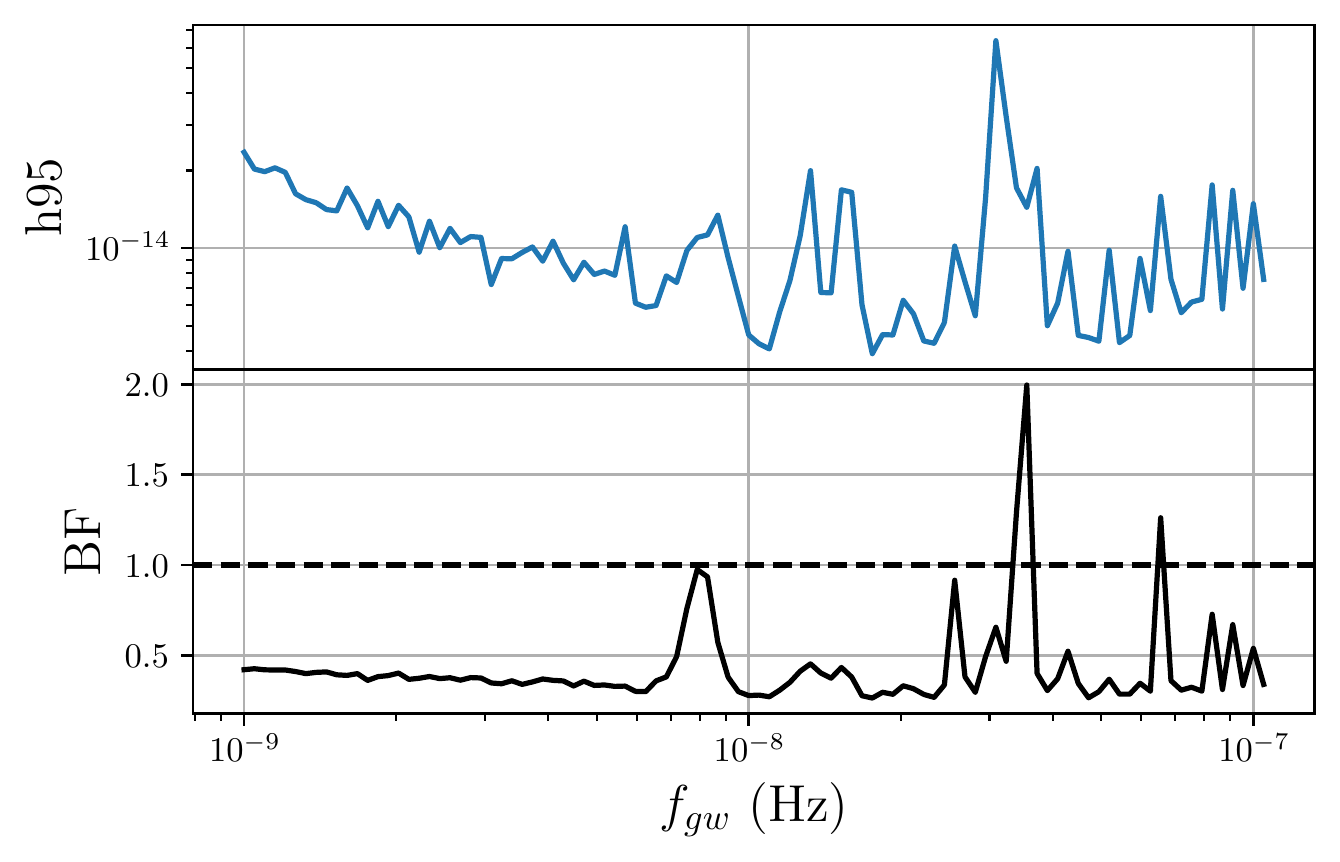}
	\caption{On the top panel, the upper bound of the 95\% central credible region on $\log h$ obtained with a log-uniform prior (with CRN). Bottom panel is the associated BF at given CGW frequency. The black dashed line shows where BF is equal to 1. The two peaks around 35 and 70 nHz with BF$>$1 are the 1yr and 2yr and should not be taken into account.}
	\label{fig:ul_bf}
\end{figure}

The main result of this paper is summarized in Figure~\ref{fig:ul_bf}. In the lower panel  we plot the BF,
and, as one can see, the noise model is usually preferred. There are few spikes where BF reaches 1, this is definitely not a detection, however PTA efforts probably should monitor carefully those frequencies in the future extended IPTA datasets. The excess in BF was also used as identification of the CGW candidates to follow.  These results show that there is nothing to follow\footnote{Note that the highest spike corresponds to frequencies close to 
$f_{yr}$ and should be discarded.}: $\mathcal{B}_{custom} ^{custom+CGW} \le 1$.

The top panel of Figure~\ref{fig:ul_bf} shows the upper bound of the 95\% central credible region for CGW strain computed at the same set of frequencies. Note the "spiky" features at several frequencies (8.1, 13, 16) nHz corresponding to the outlier  indicating potential candidates of CGW. However the corresponding BFs are (0.95, 0.33, 0.34) indicating no statistical significance in the analyzed data. 

\begin{figure}[h]
	\centering
	\includegraphics[width=0.5\textwidth]{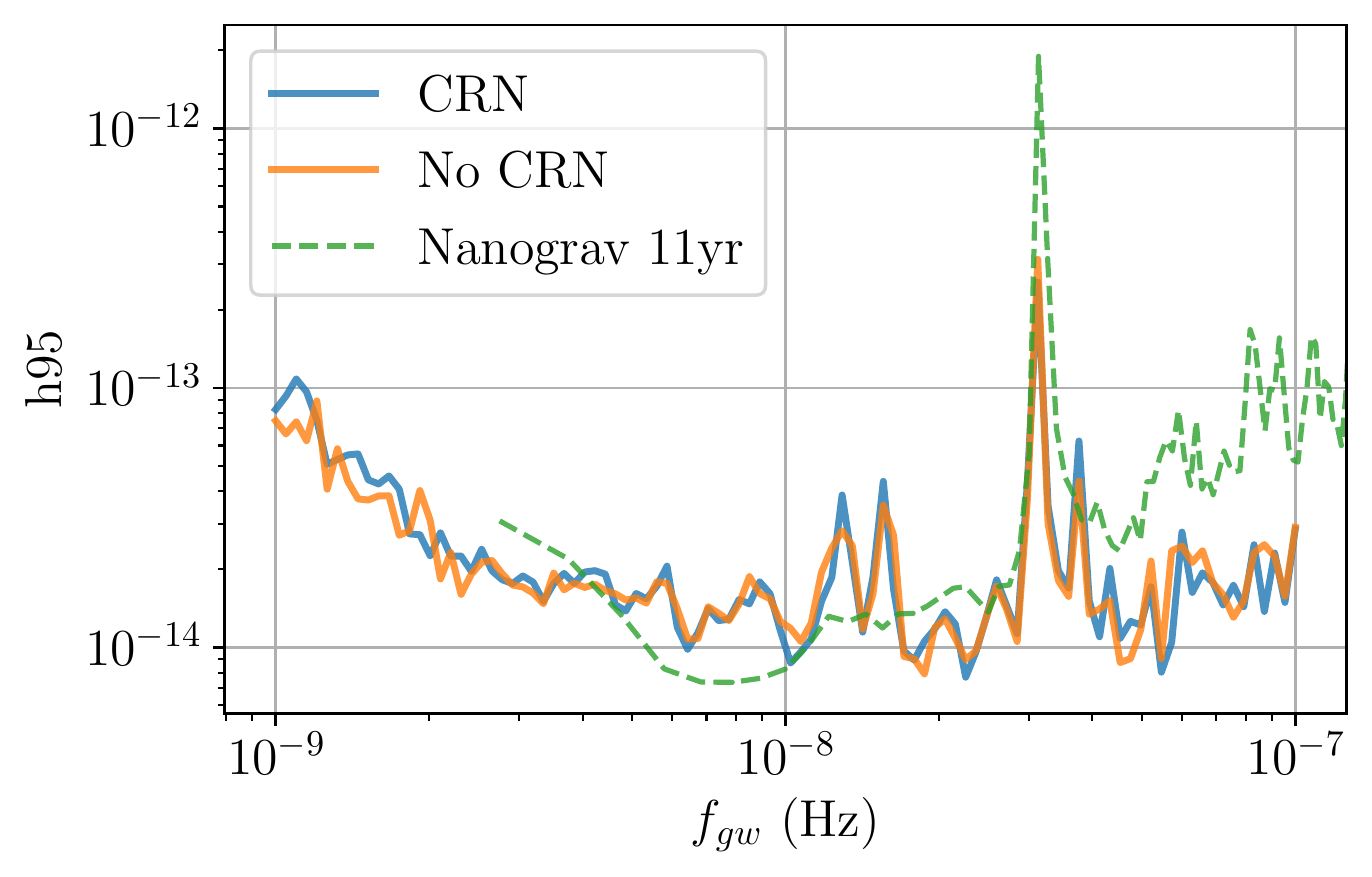}
	\caption{95\% sky averaged upper limit $h_{95}$ on CGW amplitude for models with and without CRN}
	\label{fig:ul_crn_vs_no_crn}
\end{figure}

We have computed the upper limit using uniform in GW strain prior with and without CRN on the fine frequency grid; the results are present in Figure~\ref{fig:ul_crn_vs_no_crn}. The upper limit slightly worsens at low frequencies when we add the CRN; this is understandable as we need a higher amplitude of CGW to get the same SNR when raising the noise floor. The most sensitive frequency of IPTA DR2 is  10.2 nHz with $h_{95} = 9.1 \times 10^{-15}$ in both cases with and without CRN component.

\begin{figure}
	\centering
	\includegraphics[width=0.5\textwidth]{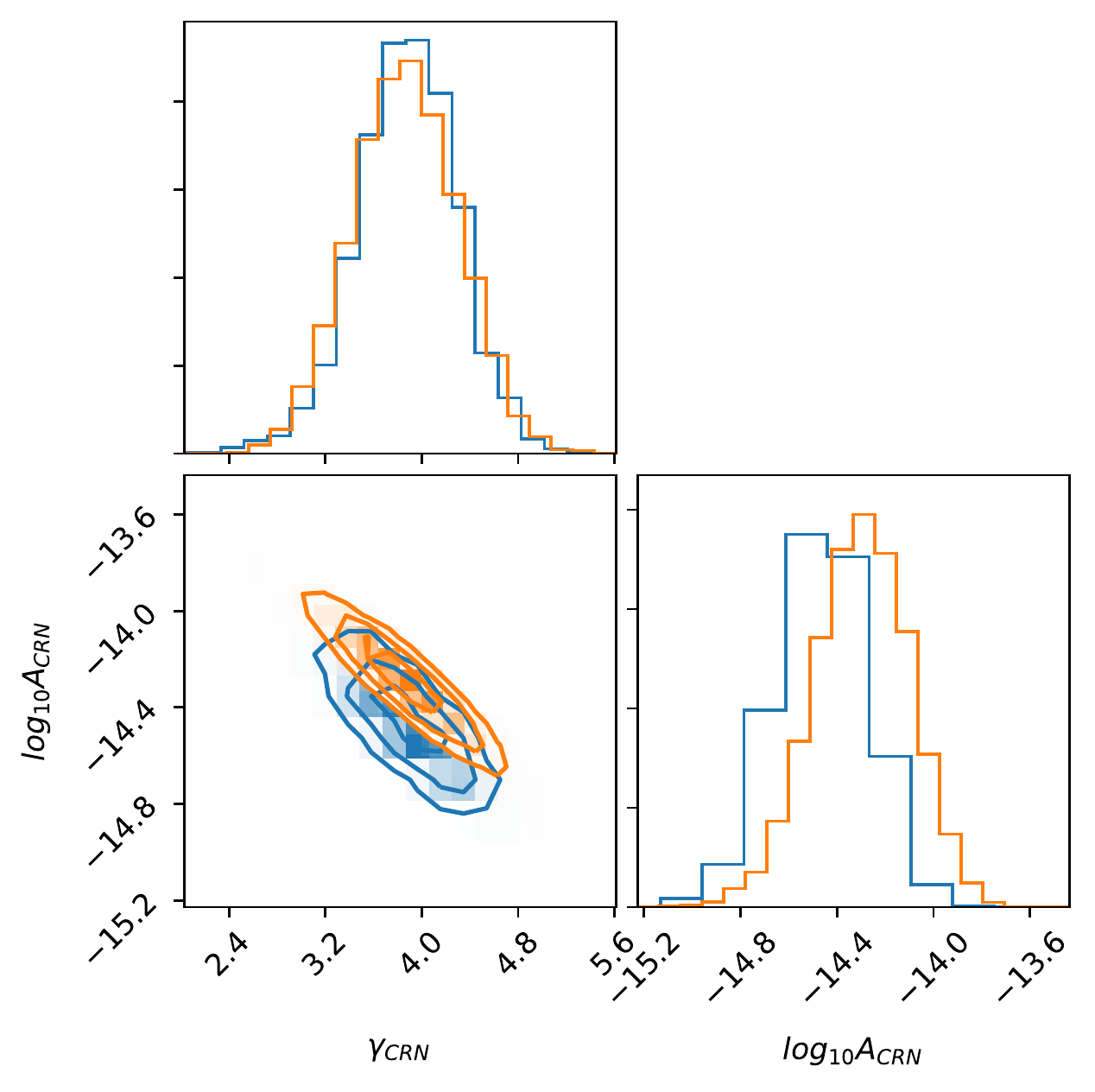}
	\caption{CRN parameters for IPTA DR2 without CGW signal obtained in (orange) and with CGW signal at 1 nHz (blue).}
	\label{fig:crn_old_vs_crn_cgw}
\end{figure}

In the model that includes CRN, we are allowing slope and amplitude to vary during the sampling. The recovered posterior for CRN with CGW at 1nHz (blue) is compared to the posterior obtained in \cite{ipta_antoniadis_2022} (orange) in figure \ref{fig:crn_old_vs_crn_cgw}.  We observe that the amplitude of CRN in the model which does not include CGW (\cite{ipta_antoniadis_2022}) is slightly higher while the slope is almost the same.  This could be explained by a partial absorption of the CRN into CGW. At the same time adding the CRN to the model increases the overall noise level (therefore decreasing the SNR of the signal). The interplay between CRN and CGW appears to mitigate the effect of the CRN on the upper limits in figure \ref{fig:crn_old_vs_crn_cgw}.

%
%

We have also overplotted the best CGW upper limit available to date based on the analysis of the  NANOGRAV 11 years dataset  \citep{nanograv_cw_11yr} as a dashed (green) line. Note that only the nine-year NANOGRAV data set was included in IPTA DR2. As expected, our current results are better at very low frequencies thanks to the longer observation time. Extended sky coverage, improved effective cadence of observation thanks to overlapping timing data (gaps coverage) and the addition of pulsars like J0437-4715 (only present in PPTA data \cite{ppta_dr1}), which is an excellent timer,  is reflected in a much improved upper limit at 100 nHz, where we might expect the first detection of CGWs.


We want to point at the double-peak feature just above 10nHz (see Figure \ref{fig:ul_bf}). This peak remains the same under the prior change (from the uniform in the amplitude to the uniform in the log-amplitude), which often corresponds to a signal present in the data. The Bayes factors for those peaks are low (0.33, 0.34). Nonetheless, we should keep an eye on those frequencies in the next data releases.

\section{Effect of noise modeling on the CGW search}
\label{sec:noise_effect}

In this subsection, we consider several noise models and compute BF between those models of noise 
with and without CGW.  The main results presented in the previous section were obtained with the custom-made noise model for the best EPTA pulsars \citep{Chalumeau_2021} and with RN30DM100 choice of Fourier frequencies for other pulsars. 
The custom noise model modifies the number of Fourier frequencies as given in table~\ref{tab:psr_fbins} 
and includes additional noise components (like system noise). However, the Fourier basis for CRN is always fixed at 30 frequency bins.

\begin{table}
	\centering
	\begin{tabular}{c|c|c|}
		Pulsar name & RN bins & DM bins \\
		\hline
		J0613-0200 & 10 & 30 \\
		J1012+5307 & 150 & 30 \\
		J1713+0747 & 15 & 150 \\
		J1744-1134 & 10 & 100 \\
		J1909-3744 & 10 & 100 \\
		J0437-4715 & 30 & 100 \\
		Other pulsars & 30 & 100 
	\end{tabular}
	\caption{Number of frequency bins used for individual RN and DM noise}
	\label{tab:psr_fbins}
\end{table}

\begin{figure}
	\centering
	\includegraphics[width=0.5\textwidth]{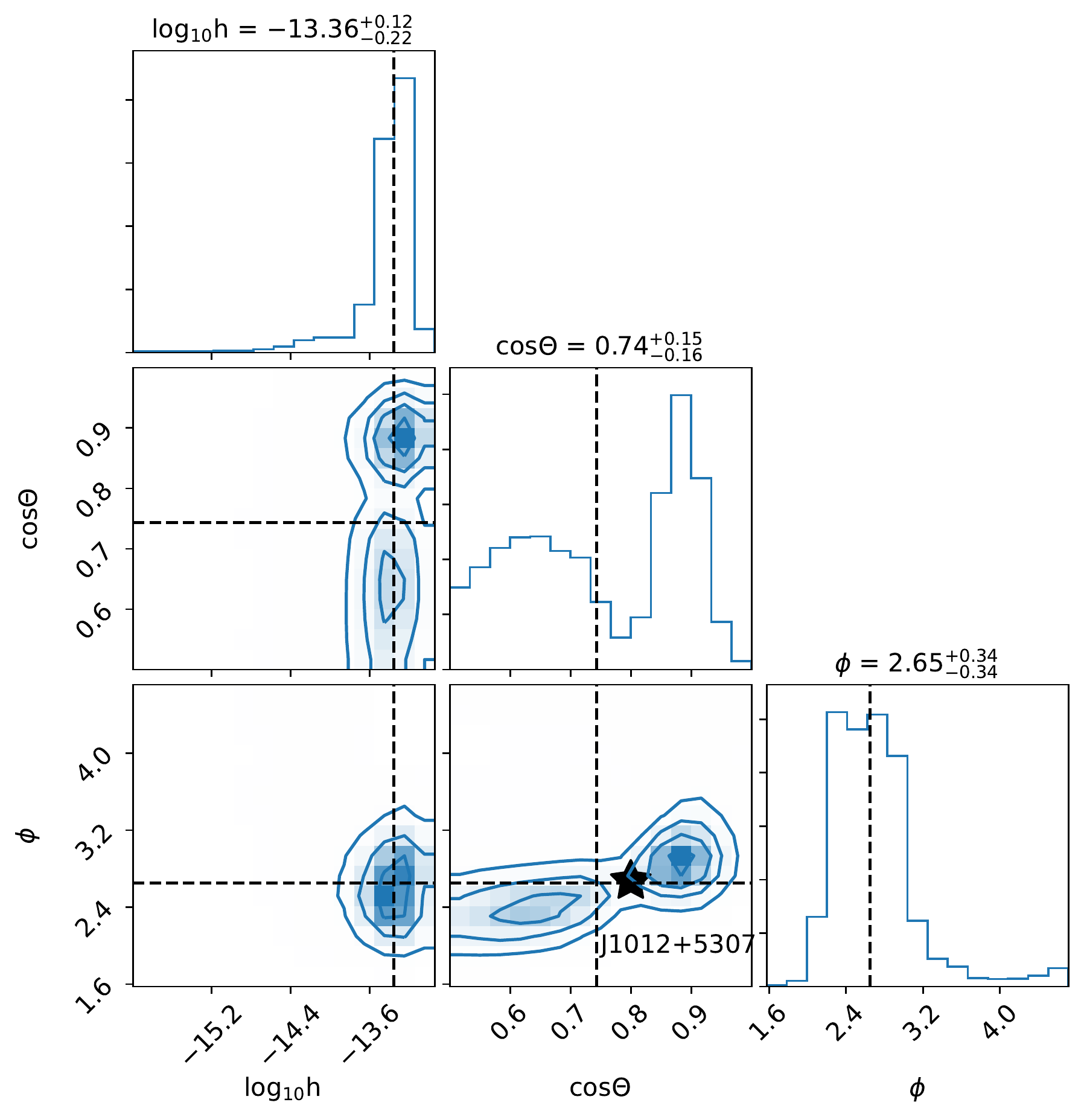}
	\caption{Posterior distribution for $\log h$, $\cos \theta$ and $\phi$ using noise model $\mathcal{M}_{RN30_DM30}$. The black star indicates the sky position of J1012+5307. The dashed lines represent the median values of the parameters. The quantile values correspond to [0.16, 0.5, 0.84].}
	\label{fig:cw_post_old}
\end{figure}

Here we present results with what was considered "standard" noise settings before this work, namely 
RN30DM30 model. We have started analysis using this model, and the search quickly converged to a particular sky position at 51  nHz.  
The first peculiarity of these results is that 51 nHz is very close to Venus orbital frequency, and the second is that the sky position had a bi-modal structure and was located very close to J1012+5307  
see Figure~\ref{fig:cw_post_old}.
The Bayes factor for this event with the RN30DM30 noise model was $\mathcal{B}_{RN30\_DM30} ^{RN30\_DM30 + CGW} = 18$, which is not very high; however, it seriously triggered our attention by being relatively well constrained in the parameter space. 

We have launched a set of investigations trying to understand this event. Using samples taken from the candidate's posterior, we have checked the contribution to the SNR from each pulsar.
 It came out that the main part comes from  J1012+5307, but a few other pulsars (J1713+0747 and J0437-4715) also contributed not negligibly. We have checked that the zero contribution from a very good timer J1909-3744 is expected given the presumed sky position of the event.
	
We also conducted  a set of injections of CGW signals with parameters taken from the candidate's posteriors using a simulated IPTA data (same TOAs as the real) with white noise only (RN and DM are supposed to be sub-dominant at the candidate frequency). We could not reproduce the observed results with injections even when we increased the amplitude of the simulated signal. More specifically, the sky location posteriors that were recovered for the injected CGW signals were not matching the double blob structure around J1012+5307 observed using the real data.

BAYESEPHEM does not include a contribution from the inner planets to the phenomenological model, so we consulted a group from Observatoire de la C\^ote d'Azur (INPOP group), inquiring if there could be an error in the Venus orbit picked up in the CGW analysis. We were reassured that the Venus parameters are known with very high precision: this is a simple planet without any moons. However, looking into the future, we probably should extend BAYESEPHEM (or alternative mitigation models) by including the perturbation of orbital elements for the inner planets. 
We have performed the analysis with a narrow prior around this event with the model including the pulsar term. The Bayes factor has slightly increased  $BF=21$; however, the parameter estimation did not change appreciably. We have tried several runs with an extended model that includes the orbital eccentricity (again with a narrow prior). The eccentric runs show a very poor convergence, but all of them suggest a relatively high eccentricity. Results of both models (eccentric and with pulsar term) imply that the power is not localized at one particular frequency but spread over some finite-size frequency band.

\begin{figure}
	\centering
	\includegraphics[width=0.5\textwidth]{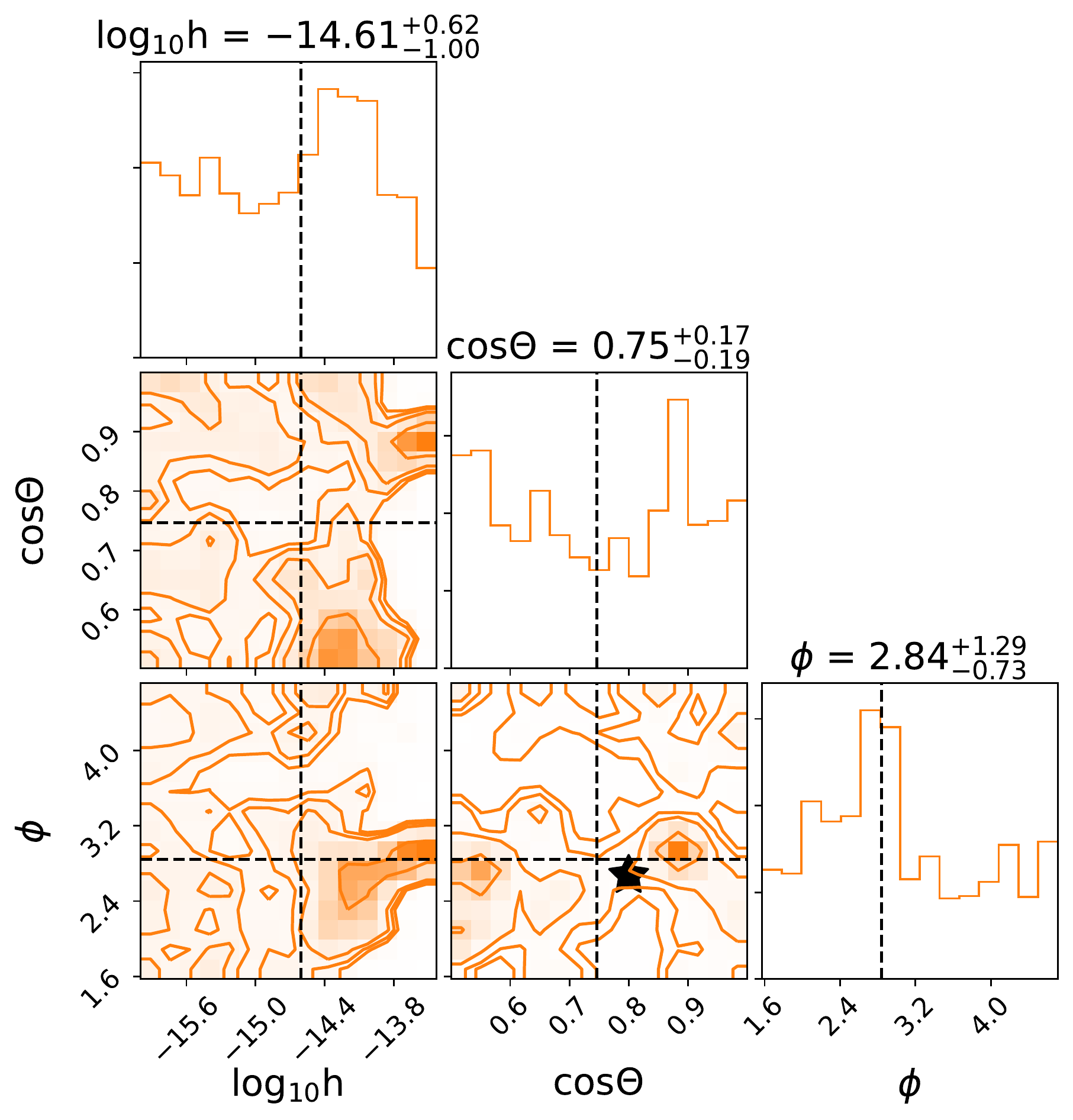}
	\caption{Posterior distribution for $\log h$, $\cos \theta$ and $\phi$ using noise model $\mathcal{M}_{custom}$. The black star indicates the sky position of J1012+5307. The dashed lines represent the median values of the parameters. The quantile values correspond to [0.16, 0.5, 0.84].}
	\label{fig:cw_post_new}
\end{figure}

We have turned to several noise models. We have started with varying number of Fourier frequencies used in the Gaussian process for RN and DM and tried $\mathcal{M}_{RN30DM100}$ and $\mathcal{M}_{RN100DM100}$. The BF for CGW with those noise models has increased tremendously (by a factor 100-1000). Finally, we have tried the custom model for the six best EPTA pulsars. Most notable is the peculiar noise model for J1012+5307 (see table~\ref{tab:psr_fbins}) which, in combination with  
the sky position of the candidate event (being next to it), suggests that the explanation might be in the time-correlated high-frequency noise present in that pulsar (see Appendix of \cite{Chalumeau_2021}).
This would also be consistent with the results of eccentric runs suggesting that this could be an extended frequency feature. We do not know the origin of that noise; it was found empirically. Including the custom noise model reduced the BF for this event 
to  $\mathcal{B}_{custom} ^{custom + CGW} = 0.95$ and the posterior samples are not anymore constrained (see Figure~\ref{fig:cw_post_new} and, please, note that we have used different range for the parameters). We have also computed the Bayes factor for the noise-only 
models (assuming that GW signal is weak) and found that 
$\mathcal{B}_{custom} ^{RN30DM30} = < 10^{-2}$ and similar result for RN30DM100 model suggesting that the data prefers by far the custom model.

Custom made noise models for PPTA pulsars were  studied in \cite{Goncharov_2020}. However, the peculiar behaviour that we have found in IPTA data was mainly originating from J1012+5307 which is not observed by PPTA. For that reason, we have chosen to focus on noise models from \cite{Chalumeau_2021}.  On the other hand, the noise model for J0437-4715 was based on \cite{Goncharov_2020} where the spectral index for RN is of 3 and the optimal number of frequency bins for DM is (at least) 91. For this pulsar, we chose 30 frequency bins for RN as it was the recommended value for spectral index > 1.5 (see \cite{Goncharov_2021}).
 
This was a useful exercise that triggered a set of investigations we would have to do in case of any CGW candidate. In addition, this section shows the importance of custom modelling noise for the best pulsars in the array, especially the noise at high frequencies, which is often partially neglected, assuming that it 
is dominated by the white (measurements) noise and it does not affect the search for the stochastic GW signal (which is most pronounced at low frequencies). The considered event shows how unmodelled high-frequency noise could conspire for CGW signal.

\section{Conclusion}
\label{sec:concl}

We have searched for a continuous GW signal in the IPTA DR2 dataset. We have used the Bayesian approach and based detection criteria on the Bayes factor. We have shown that using a custom noise model for the six best EPTA pulsars is essential for the correct interpretation of the data. This is especially true for J1012+5307, which 
exhibits time-correlated noise at high frequencies. We found no CGW candidates
using this noise model and proceeded to set the upper limit on GW strain. The addition of CRN in the noise model slightly affects the upper limit by lowering the sensitivity of the array at low frequencies. The most sensitive frequency appears to be around 10 nHz with a 95\% sky averaged upper limit for CGW amplitude $h_{95} = 9.1 \times 10^{-15}$. The IPTA  DR2 shows a much better upper limit than previously set at higher frequencies, making it a promising dataset to detect CGW. 

During the analysis, we demonstrated the CGW candidate follow-up investigations program, which was an important exercise that should be used in subsequent PTA CGW analysis. The expected CGW signal has low SNR, and its SNR will be slowly accumulated as we get more pulsars and a longer observational span. Modelling noise in pulsar data is essential, especially at high frequencies. 



This analysis was limited to circular SMBHBs using only the Earth term. The use of eccentric CGW signal and including the pulsar term might potentially improve the search; however, it brings signal complexity which might make harder the interpretation of the results and increases the parameter space. We are entering the era of very high quality and high cadence radio observations with new instruments like FAST \citep{FAST} or SKA \citep{SKA} with sophisticated data analysis techniques. Additional investigations of the best approach to detecting CGW have to be re-investigated, probably using simulated data and/or an extended CGW signal injection campaign.

\section{Acknowledgments}

The International Pulsar Timing Array (IPTA) is a consortium of existing Pulsar Timing Array collaborations, namely, the European Pulsar Timing Array (EPTA), North American Nanohertz Observatory for Gravitational Waves (NANOGrav), Parkes Pulsar Timing Array (PPTA) and the recent addition of the Indian Pulsar Timing Array (InPTA). Observing collaborations from China and South Africa are also part of the IPTA.

The EPTA is a collaboration between European and partner institutes with the aim to provide high precision pulsar timing to work towards the direct detection of low-frequency gravitational waves. An Advanced Grant of the European Research Council to implement the Large European Array for Pulsars (LEAP) also provides funding. Part of this work is based on observations with the 100-m telescope of the Max-Planck-Institut f\"ur Radioastronomie (MPIfR) at Effelsberg in Germany. Pulsar research at the Jodrell Bank Centre for Astrophysics and the observations using the Lovell Telescope are supported by a Consolidated Grant (ST/T000414/1) from the UK's Science and Technology Facilities Council. The Nan\c cay radio Observatory is operated by the Paris Observatory, associated to the French Centre National de la Recherche Scientifique (CNRS), and to the Universit{\'e} d' Orl{\'e}ans. We acknowledge financial support from "Programme National de Cosmologie and Galaxies" (PNCG), and "Programme National Hautes Energies" (PNHE) funded by CNRS/INSUIN2P3-INP, CEA and CNES, France. We acknowledge financial support from Agence Nationale de la Recherche (ANR-18-CE31-0015), France. The Westerbork Synthesis Radio Telescope is operated by the Netherlands Institute for Radio Astronomy (ASTRON) with support from the Netherlands Foundation for Scientific Research (NWO). The Sardinia Radio Telescope (SRT) is funded by the Department of University and Research (MIUR), the Italian Space Agency (ASI), and the Autonomous Region of Sardinia (RAS) and is operated as National Facility by the National Institute for Astrophysics (INAF).

The NANOGrav Physics Frontiers Center is supported by the National Science Foundation (NSF) Physics Frontier Center award numbers 1430284 and 2020265. The National Radio Astronomy Observatory is a facility of the National Science Foundation operated under cooperative agreement by Associated Universities, Inc. The Green Bank Observatory is a facility of the National Science Foundation operated under cooperative agreement by Associated Universities, Inc. The Arecibo Observatory is a facility of the National Science Foundation operated under cooperative agreement by the University of Central Florida in alliance with Yang Enterprises, Inc. and Universidad Metropolitana.

The Parkes radio telescope (Murriyang) is part of the Australia Telescope which is funded by the Commonwealth Government for operation as a National Facility managed by CSIRO.

JA acknowledges support by the Stavros Niarchos Foundation (SNF) and the Hellenic Foundation for Research and Innovation (H.F.R.I.) under the 2nd Call of "Science and Society" Action Always strive for excellence - "Theodoros Papazoglou" (Project Number: 01431). SBS acknowledges generous support by the NSF through grant AST-1815664. The work is supported by National SKA program of China 2020SKA0120100, Max-Planck Partner Group, NSFC 11690024, CAS Cultivation Project for FAST Scientific. JACC was supported in part by NASA CT Space Grant PTE Federal Award Number 80NSSC20M0129. CMFM and JACC are also supported by the National Science Foundations NANOGrav Physics Frontier Center, Award Number 2020265. The Center for Computational Astrophysics is a division of the Flatiron Institute in New York City, which is supported by the Simons Foundation. This research was supported in part by the National Science Foundation grant AST-2106552. AC acknowledges support from the Paris \^Ile-de-France Region. Support for HTC was provided by NASA through the NASA Hubble Fellowship Program grant HST-HF2-51453.001 awarded by the Space Telescope Science Institute, which is operated by the Association of Universities for Research in Astronomy, Inc., for NASA, under contract NAS5-26555. SD is the recipient of an Australian ResearchCouncil Discovery Early Career Award (DE210101738) funded by the Australian Government. GD, RK and MKr acknowledge support from European Research Council (ERC) Synergy Grant "BlackHoleCam" Grant Agreement Number 610058 and ERC Advanced Grant "LEAP" Grant Agreement Number 337062. TD and MTL  acknowledge support received from NSF AAG award number 200968. ECF is supported by NASA under award number 80GSFC17M0002.002. BG is supported by the Italian Ministry of Education, University and Research within the PRIN 2017 Research Program Framework, n. 2017SYRTCN. Portions of this work performed at the Naval Research Laboratory is supported by NASA and ONR 6.1 basic research funding. Part of this research was carried out at the Jet Propulsion Laboratory, California Institute of Technology, under a contract with the National Aeronautics and Space Administration. JWM gratefully acknowledges support by the Natural Sciences and Engineering Research Council of Canada (NSERC), [funding reference CITA 490888-16]. KDO was supported in part by NSF Grant No. 2207267. ASa, ASe and GS acknowledge financial support provided under the European Union's H2020 ERC Consolidator Grant "Binary Massive Black Hole Astrophysic" (B Massive, Grant Agreement: 818691). RMS acknowledges support through Australian Research Council Future Fellowship FT190100155. JJS is supported by an NSF Astronomy and Astrophysics Postdoctoral Fellowship under award AST-2202388, and this research was supported in part by NSF AST-1847938. This research was funded partially by the Australian Government through the Australian Research Council (ARC), grants CE170100004 (OzGrav) and FL150100148. Pulsar research at UBC is supported by an NSERC Discovery Grant and by the Canadian Institute for Advanced Research. SRT acknowledges support from NSF grants AST-2007993 and PHY-2020265. SRT also acknowledges support from the Vanderbilt University College of Arts \& Science Dean's Faculty Fellowship program. Multiple NANOGrav members acknowledge support of NSF Physics Frontiers Center awards 1430284 and 2020265. SMR is a CIFAR Fellow. AV acknowledges the support of the Royal Society and Wolfson Foundation. JPWV acknowledges support by the Deutsche Forschungsgemeinschaft (DFG) through the Heisenberg programme (Project No. 433075039). MED acknowledges support from the National Science Foundation (NSF) Physics Frontier Center award 1430284, and from the Naval Research Laboratory by NASA under contract S-15633Y. ZCC is supported by the National Natural Science Foundation of China (Grant No. 12247176) and the China Postdoctoral Science Foundation Fellowship No. 2022M710429. CAW acknowledges support from CIERA, the Adler Planetarium, and the Brinson Foundation through a CIERA-Adler postdoctoral fellowship

\section*{Author contribution}

This paper's author list follows an alphabetically ordered two tier structure in which the first tier lists the main contributors to this work and the second lists all other members of the collaboration.

All analyses and plots were produced by M. Falxa with S. Babak as the main adviser. AC, APe, SJV contributed to the production of the pulsar noise models and to setting up the main data analysis scripts. PTB, BB, PRB, SC, LG, JSH, CMFM, APa, APe, NSP, AS SBS, SRT, GT, CAW provided useful guidance and suggested additional analyses that improved the quality of this work.  Additionally BB, ZC, NC, MV, SJV, CAW, XZ have been involved in searches for continuous gravitational waves using other PTA datasets, and we have adopted (where appropriate) their analysis code, methods and used past publications as guidelines for this paper.

The additional authors have contributed to this work through the radio observations, production of the data, and the combination of datasets from individual PTAs into IPTA DR2. For more details on individual roles, please refer to \cite{ipta_antoniadis_2022}.

\section*{Data availability}

The timing data used in this article is available on the IPTA website \url{https://ipta4gw.org/data-release/} (second data release : \url{https://gitlab.com/IPTA/DR2}).

\bibliographystyle{mnras}
\bibliography{biblio_ipta_cw.bib}



\section*{Affiliations}
	\small \it{
		$^{1}$ Universit{\'e} de Paris, CNRS, Astroparticule et Cosmologie, 75013 Paris, France\\
		$^{2}$ Department of Physics and Astronomy, Widener University, One University Place, Chester, PA 19013, USA\\
		$^{3}$ Department of Physics, Montana State University, Bozeman, MT 59717, USA\\
		$^{4}$ Dipartimento di Fisica ``G. Occhialini", Universit{\'a} degli Studi di Milano-Bicocca, Piazza della Scienza 3, 20126 Milano, Italy\\
		$^{5}$ Kavli Institute for Astronomy and Astrophysics, Peking University, Beijing 100871, P. R. China\\
		$^{6}$ Department of Astronomy, Beijing Normal University, Beijing 100875, China\\
		$^{7}$ Station de Radioastronomie de Nan\c{c}ay, Observatoire de Paris, PSL University, CNRS, Universit{\'e} d'Orl{\'e}ans, 18330 Nan\c{c}ay, France\\
		$^{8}$ Laboratoire de Physique et Chimie de l'Environnement et de l'Espace LPC2E UMR7328, Universit{\'e} d'Orl{\'e}ans, CNRS, 45071 Orl{\'e}ans, France\\
		$^{9}$ Department of Physics, Oregon State University, Corvallis, OR 97331, USA\\
		$^{10}$ Center for Computational Astrophysics, Flatiron Institute, 162 5th Avenue, New York, NY 10010, USA\\
		$^{11}$ Department of Physics, University of Connecticut, 196 Auditorium Road, U-3046, Storrs, CT 06269-3046, USA\\
		$^{12}$ Max-Planck-Institut f{\"u}r Radioastronomie, Auf dem H{\"u}gel 69, 53121 Bonn, Germany\\
		$^{13}$ IRFU, CEA, Universit\'e Paris-Saclay, F-91191, Gif-sur-Yvette, France\\
		$^{14}$ Department of Physics and Astronomy, Vanderbilt University, 2301 Vanderbilt Place, Nashville, TN 37235, USA\\
		$^{15}$ INFN, Sezione di Milano-Bicocca, Piazza della Scienza 3, 20126 Milano, Italy\\
		$^{16}$ Department of Physics and Astronomy, West Virginia University, P.O. Box 6315, Morgantown, WV 26506, USA\\
		$^{17}$ Laboratoire Univers et Th{\'e}ories LUTh, Observatoire de Paris, Universit{\'e} PSL, CNRS, Universit{\'e} de Paris, 92190 Meudon, France\\
		$^{18}$ Jet Propulsion Laboratory, California Institute of Technology, 4800 Oak Grove Drive, Pasadena, CA 91109, USA\\
		$^{19}$ Center for Gravitation, Cosmology and Astrophysics, Department of Physics, University of Wisconsin-Milwaukee, P.O. Box 413, Milwaukee, WI 53201, USA\\
		$^{20}$ Center for Interdisciplinary Exploration and Research in Astrophysics (CIERA), Northwestern University, Evanston, IL 60208, USA\\
		$^{21}$ Adler Planetarium, 1300 S. DuSable Lake Shore Dr., Chicago, IL 60605, USA\\
		$^{22}$ Advanced Institute of Natural Sciences, Beijing Normal University, Zhuhai 519087, China\\
		$^{23}$ Institute of Astrophysics, FORTH, N. Plastira 100, 70013, Heraklion, Greece\\
		$^{24}$ X-Ray Astrophysics Laboratory, NASA Goddard Space Flight Center, Code 662, Greenbelt, MD 20771, USA\\
		$^{25}$ Centre for Astrophysics and Supercomputing, Swinburne University of Technology, P.O. Box 218, Hawthorn, Victoria 3122, Australia, ARC Centre of Excellence for Gravitational Wave Discovery (OzGrav)\\
		$^{26}$ International Centre for Radio Astronomy Research, Curtin University, Bentley, WA 6102, Australia\\
		$^{27}$ Department of Physics, University of Florida, 2001 Museum Rd., Gainesville, FL 32611-8440, USA\\
		$^{28}$ Cornell Center for Astrophysics and Planetary Science and Department of Astronomy, Cornell University, Ithaca, NY 14853, USA\\
		$^{29}$ Centre for Astrophysics and Supercomputing, Swinburne University of Technology, P.O. Box 218, Hawthorn, Victoria 3122, Australia\\
		$^{30}$ Department of Physics and Astronomy, Franklin \& Marshall College, P.O. Box 3003, Lancaster, PA 17604, USA\\
		$^{31}$ Department of Physics and Astronomy, University of British Columbia, 6224 Agricultural Road, Vancouver, BC V6T 1Z1, Canada\\
		$^{32}$ School of Science, Western Sydney University, Locked Bag 1797, Penrith South DC, NSW 2751, Australia\\
		$^{33}$ George Mason University, resident at the Naval Research Laboratory, Washington, DC 20375, USA\\
		$^{34}$ National Radio Astronomy Observatory, 1003 Lopezville Rd., Socorro, NM 87801, USA\\
		$^{35}$ Department of Physics, Hillsdale College, 33 E. College Street, Hillsdale, MI 49242, USA\\
		$^{36}$ Eureka Scientific, 2452 Delmer Street, Suite 100, Oakland, CA 94602-3017, USA\\
		$^{37}$ Laboratory for Multiwavelength Astrophysics, Rochester Institute of Technology, Rochester, NY 14623, USA\\
		$^{38}$ Research Center for Intelligent Computing Platforms, Zhejiang Laboratory, Hangzhou 311100, China\\
		$^{39}$ Department of Astronomy, University of Maryland, College Park, MD 20742, USA\\
		$^{40}$ Department of Physics and Astronomy, West Virginia University, PO Box 6315, Morgantown, WV 26506, USA\\
		$^{41}$ WVU Center for Gravitational Waves and Cosmology, White Hall, Morgantown, WV 26506 , USA\\
		$^{42}$ Gran Sasso Science Institute, Viale Francesco Crispi, 7, 67100 L'Aquila AQ, Italy\\
		$^{43}$ Department of Astronomy, University of Michigan, 1085 S. University Ave, Ann Arbor, MI 48109, USA\\
		$^{44}$ CSIRO Space and Astronomy, Australia Telescope National Facility, PO Box 76, Epping NSW 1710, Australia\\
		$^{45}$ Theoretical AstroPhysics Including Relativity (TAPIR), MC 350-17, California Institute of Technology, Pasadena, CA 91125, USA\\
		$^{46}$ Dominion Radio Astrophysical Observatory, Herzberg Research Centre for Astronomy and Astrophysics, National 27 Research Council Canada, PO Box 248, Penticton, BC V2A 6J9, Canada\\
		$^{47}$ Jodrell Bank Centre for Astrophysics, Department of Physics and Astronomy, University of Manchester, Manchester M13 9PL, UK\\
		$^{48}$ Space Science Division, Naval Research Laboratory, Washington, DC 20375-5352, USA\\
		$^{49}$ University of Washington Bothell, 18115 Campus Way NE, Bothell, WA 98011, USA\\
		$^{50}$ Oregon State University, Department of Physics, 301 Weniger Hall, Corvallis, OR, 97331, USA\\
		$^{51}$ Department of Astronomy \& Astrophysics, University of Toronto, 50 Saint George Street, Toronto, ON M5S 3H4, Canada\\
		$^{52}$ Green Bank Observatory, P.O. Box 2, Green Bank, WV 24944, USA\\
		$^{53}$ Department of Physics, University of the Pacific, 3601 Pacific Ave., Stockton, CA 95211, USA\\
		$^{54}$ E.A. Milne Centre for Astrophysics, University of Hull, Cottingham Road, Kingston-upon-Hull, HU6 7RX, UK\\
		$^{55}$ Centre of Excellence for Data Science, Artificial Intelligence and Modelling (DAIM), University of Hull, Cottingham Road, Kingston-upon-Hull, HU6 7RX, UK\\
		$^{56}$ Dunlap Institute for Astronomy and Astrophysics, University of Toronto, 50 Saint George Street, Toronto, ON M5S 3H4, Canada\\
		$^{57}$ Department of Physics, Lafayette College, Easton, PA 18042, USA\\
		$^{58}$ Institute of Cosmology, Department of Physics and Astronomy, Tufts University, Medford, MA 02155, USA\\
		$^{59}$ Manly Astrophysics, 15/41-42 East Esplanade, Manly, NSW 2095, Australia\\
		$^{60}$ Institute of Physics, E\"otv\"os Lor\'and University, P\'azm\'any P.s. 1/A, 1117 Budapest, Hungary\\
		$^{61}$ Arecibo Observatory, HC3 Box 53995, Arecibo, PR 00612, USA\\
		$^{62}$ INAF - Osservatorio Astronomico di Cagliari, via della Scienza 5, 09047 Selargius (CA), Italy\\
		$^{63}$ National Radio Astronomy Observatory, 520 Edgemont Road, Charlottesville, VA 22903, USA\\
		$^{64}$ U.S. Naval Research Laboratory, Washington, DC 20375, USA\\
		$^{65}$ CSIRO Scientific Computing, Australian Technology Park, Locked Bag 9013, Alexandria, NSW 1435, Australia\\
		$^{66}$ CSIRO Parkes Observatory, PO Box 276, Parkes NSW 2870, Australia\\
		$^{67}$ Giant Army, 915A 17th Ave, Seattle, WA, 98122, USA\\
		$^{68}$ Department of Astrophysical and Planetary Sciences, University of Colorado, Boulder, CO 80309, USA\\
		$^{69}$ Center for Astrophysics | Harvard \& Smithsonian\\
		$^{70}$ Department of Physics and Astronomy, Swarthmore College, Swarthmore, PA 19081, USA\\
		$^{71}$ Max Planck Institute for Gravitational Physics (Albert Einstein Institute), Am Mu{\"u}hlenberg 1, 14476 Potsdam, Germany\\
		$^{72}$ Department of Physics and Astronomy, Oberlin College, Oberlin, OH 44074, USA\\
		$^{73}$ Institute for Gravitational Wave Astronomy and School of Physics and Astronomy, University of Birmingham, Edgbaston, Birmingham B15 2TT, UK\\
		$^{74}$ Fakult{\"a}t f{\"u}r Physik, Universit{\"a}t Bielefeld, Postfach 100131, 33501 Bielefeld, Germany\\
		$^{75}$ Xinjiang Astronomical Observatory, Chinese Academy of Sciences\\
		$^{76}$ Institute of Optoelectronic Technology, Lishui University, Lishui, 323000, China\\
		$^{77}$ National Astronomical Observatories, Chinese Academy of Sciences, Beijing 100101, China\\
		$^{78}$ Purple Mountain Observatory, Chinese Academy of Sciences, Nanjing 210023, China\\
}

\bsp	
\label{lastpage}

\end{document}